\documentclass[runningheads,a4paper]{llncs}

\usepackage{csquotes}
 
\usepackage{amssymb}
\usepackage{amsmath,mathtools}

\usepackage{graphicx}

\usepackage[percent]{overpic}
\usepackage{color}

\usepackage{url}
\urldef{\mailsa}\path|{elenasi, algh, iolanda, celle, dani}@kth.se|
  
\newcommand{\keywords}[1]{\par\addvspace\baselineskip
\noindent\keywordname\enspace\ignorespaces#1}

\begin{document}

\mainmatter  

\title{Exploring Temporal Dependencies\\in Multimodal Referring Expressions\\with Mixed Reality}

\titlerunning{Exploring Temporal Dependencies in Multimodal Referring Expressions}

%
%
\author{Elena Sibirtseva$^{1}$
\and Ali Ghadirzadeh$^{1,2}$\and Iolanda Leite$^{1}$ \and\\M\r{a}rten Bj\"{o}rkman$^{1}$\and
Danica Kragic$^{1}$}
\authorrunning{Exploring Temporal Dependencies in Multimodal Refferring Expressions}

\institute{$^{1}$ Devision of Robotics, Perception and Learning ,\\EECS at KTH Royal Institute of Technology, Stockholm, Sweden\\
$^{2}$ Intelligent Robotics Research Group,\\
Aalto University, Espoo, Finland\\
\mailsa\\
}

%
%

\toctitle{}
\tocauthor{}
\maketitle

\begin{abstract}
In collaborative tasks, people rely both on verbal and non-verbal cues simultaneously to communicate with each other. 
For human-robot interaction to run smoothly and naturally, a robot should be equipped with the ability to robustly disambiguate referring expressions. 
In this work, we propose a model that can disambiguate multimodal fetching requests using modalities such as head movements, hand gestures, and speech. We analysed the acquired data from mixed reality experiments and formulated a hypothesis that modelling temporal dependencies of events in these three modalities increases the model's predictive power. We evaluated our model on a Bayesian framework to interpret referring expressions  
with and without exploiting the temporal prior.
\end{abstract}

\section{Introduction}

In most industrial applications, robots typically work in isolation from humans in repetitive tasks with or without very little interaction with humans. There has been however, the need for developing collaborative robots that can communicate their intent to humans, but also understand human communicative behaviours \cite{thomaz2016computational}.
This means that we need to go beyond designing classical pre-scripted robots for industrial settings, and more towards assistive robot co-workers with interaction capabilities that empower human workers. For humans to establish successful communication with robotic agents, robots need to use multisensory approaches to perceive human multimodal data and \textit{interpret social cues that communicate humans' intent}. 

One of the challenges in understanding human intent is the multisensory fusion problem \cite{turk2014multimodal,lalanne2009fusion}. The goal of multisensory fusion is to get data from different sensors, combine it in some fashion and, ultimately, a come up with a model of the user's intentions. The main problem is interpretation of each modality in combination with each other. Our focus in this research is currently on the following modalities: head movements, hand gestures and speech. The interaction scenario we are interested in is a fetching task, where a human participant explains which object he/she needs from the shared workspace and the robot has to interpret the request from the multimodal sensor observations.

\begin{figure}
\centering
\includegraphics[width=0.5\columnwidth,scale=0.5]{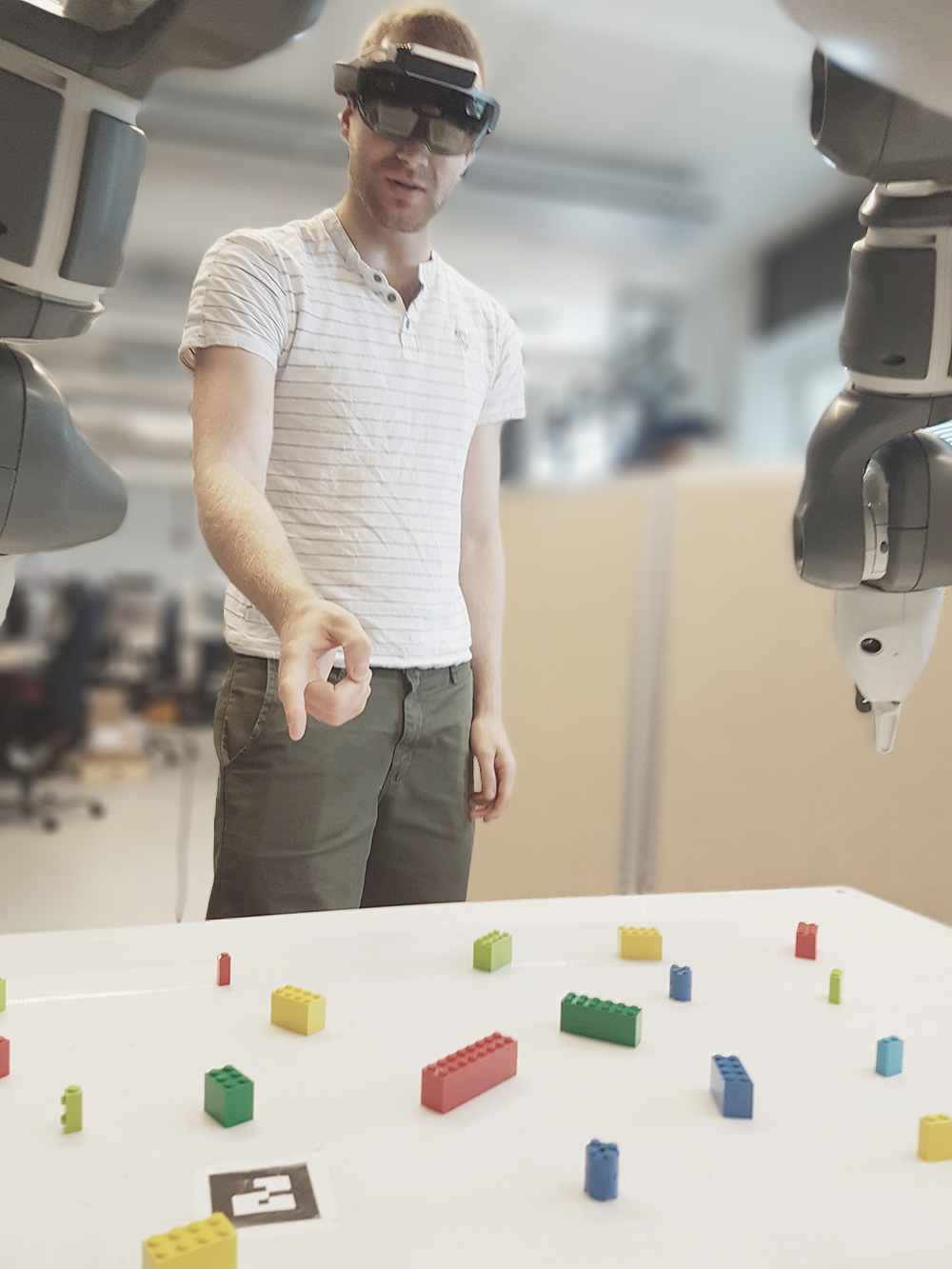}
\caption{Interaction scenario. The human agent is requesting a robot to give them Lego blocks. Any modality can be used in a natural way, no restriction to the interaction is applied. The only tracked modalities are head movements, hand gestures and speech. The participant wears a mixed reality headset to see which block to request and what is the robot's estimation of their command. }
\label{fig:interaction}
\end{figure}

Recent studies focused on intent recognition by combining different features from speech with gaze fixations \cite{admoni2016predicting}, head movements \cite{veronese2017probabilistic}, and gestures \cite{duarte2018action}. However, in non-guided natural human-robot interaction this approach has its own limitations. Our previous human study \cite{kontogiorgos2018multimodal} showed that participants often look at each other more than at the target object or spend more time looking at the next object in the sequence while still describing the previous one. This lead us to look for more high level behaviour patterns that consist of events happening in all the modalities. Our hypothesis is that it is important to look at \textit{when} a certain event happened (e.g. head fixation, pointing gesture) given events in other modalities and \textit{not for how long}. This way we assume that individual events in modalities can be combined in higher level behaviour patterns based on temporal dependencies.

In this paper we present our findings on the answers to the following questions: 
\begin{itemize}
\item \textbf{Q1:} Do common temporal patterns emerge in participants' behaviour during the fetching request task? 
\item \textbf{Q2:} If we encode these patterns as temporal priors, will they be helpful in inferring the intended object from multimodal referring expressions? 
\item \textbf{Q3:} Are temporal patterns common across most participants or are they person-dependent? 
\end{itemize}

We discuss what we learned from the analysis of a human study and how we see the future development of efficient and natural human-robot interaction in shared workspaces.

\section{Related Work}

Disambiguation of referring expressions is a well-researched topic in the human-robot interaction community. While written text understanding can be performed in batches, real-time interaction requires continuous reference resolution.  

Many works focused exclusively on the language part of the request through incremental reference resolution \cite{chai2014collaborative,funakoshi2012unified}. However incremental reference resolution is sometimes not enough to completely disambiguate a verbal request. Additional information can be inferred from other modalites, since studies show that people convey considerable amount of information through non-verbal cues \cite{thomaz2016computational}. For instance, through gaze \cite{mehlmann2014exploring,admoni2016predicting}, head movements \cite{veronese2017probabilistic}, and gestures \cite{duarte2018action}. Our focus is on combining three modalities - speech, head movements, and pointing gestures.  

While originally relationships between modalities were encoded through a heuristic approach \cite{bolt1980put}, currently probabilistic graphical models \cite{savran2015temporal,srivastava2012learning} and deep learning \cite{venugopalan2014translating,yao2015describing} are the most common ways to handle the multimodal representation. We are interested in investigating multimodal behaviour patters and modelling them explicitly in a probabilistic manner. Thus we implemented multimodal fusion as a Bayesian filter, which already showed promising results for reference resolution \cite{whitney2016interpreting}.

More specifically, Whitney et al. \cite{whitney2016interpreting} developed a Bayesian filter to calculate the belief of an object being the target given observed person's gestures and speech. In this approach, the longer a person is pointing at an object, the more probable it is to be the target. Basing prediction on the longest fixation in continuous modalities such as pointing and gaze \cite{huang2015using} are a common way to model them. However, as was shown in our previous study \cite{kontogiorgos2018multimodal}, when the complexity of the task is increased, nosiness of these modalities increases accordingly to such an extent that it becomes nearly impossible to make a prediction solely from the longest fixation. 

Behaviour studies \cite{bavelas2014hand} showed that gestures are temporally coordinated with speech, when people are retelling a scene from their favourite movie. Based on these finding, we hypothesise that by incorporating timing of gestures and head fixations with relation to speech in our model, we can filter out unrelated to the request non-verbal behaviour. In our work, we expand Whitney's framework \cite{whitney2016interpreting} by learning a temporal prior and adding it to the observations update. Our focus is on the temporal relationships between events in modalities and whether this prior can increase filter's accuracy.


\section{Methodology}

\begin{figure}
\centering
\includegraphics[width=1\columnwidth,scale=0.5]{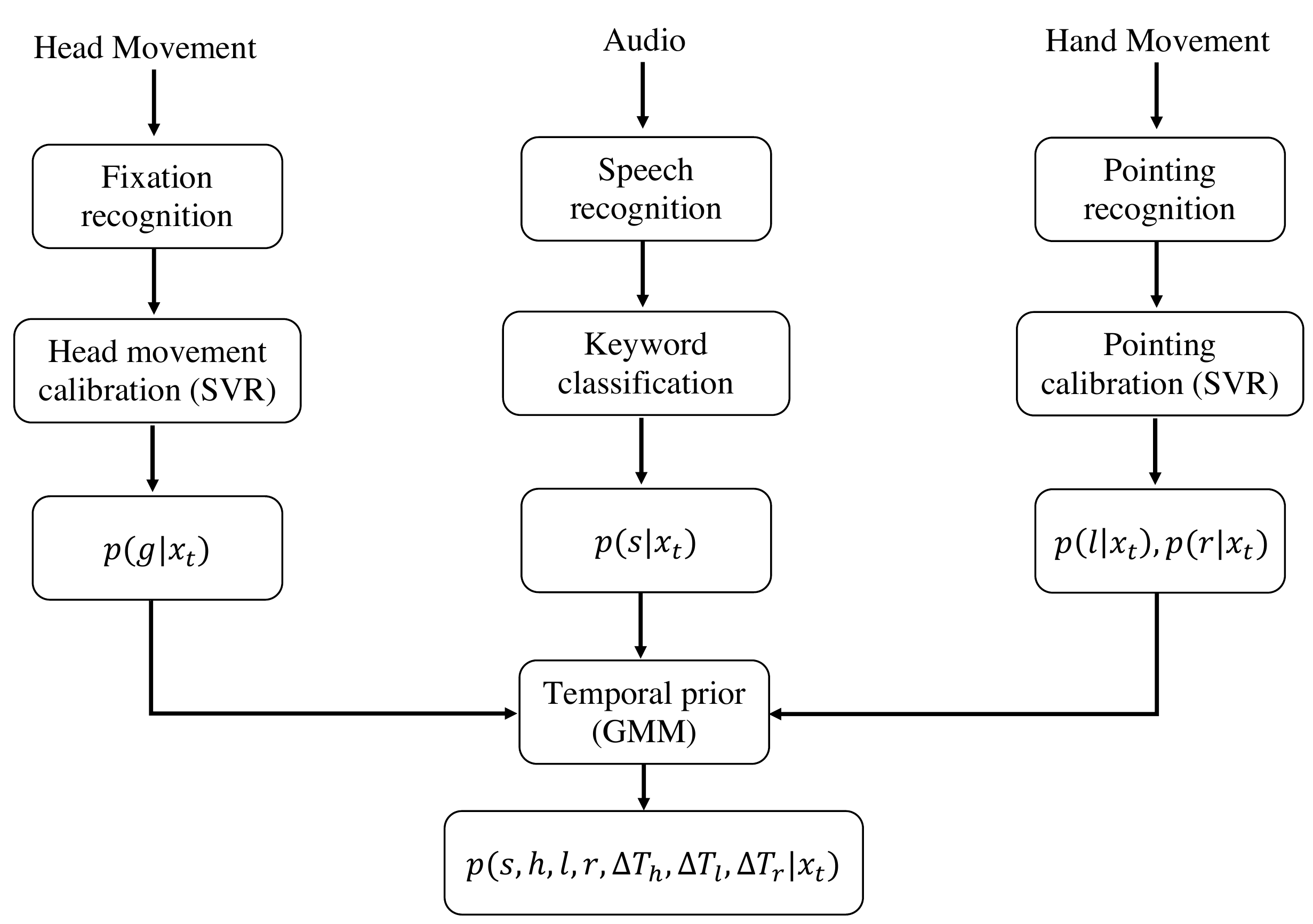}
\caption{Observation update with a temporal prior}
\label{fig:observation}
\end{figure}

\subsection{Bayesian Filter}

Having three continuous modalities, we want to fuse them together in order to get a probability distribution over all objects ($\mathcal{X}$), given observed speech, head, and pointing inputs ($\mathcal{Z}$) at each time step ($t$). We apply probabilistic inference based on Bayesian filtering \cite{thrun2005probabilistic}. The hidden state, $x_t \in \mathcal{X}$, is the target object in the scene that the person is currently referencing. The robot observes the user's non-verbal actions and speech, $\mathcal{Z}$, and at each time step estimates a distribution over the current state, $x_t$:

\begin{equation}
p(x_t|z_{0:t}).
\end{equation}

First, a prediction about current state is made based only on previous observations and then two types of update are made: time update without any contextual information and observation update, as proposed in \cite{whitney2016interpreting}.

\subsection{Observation update}

The posterior distribution of $x_t$ given a history of observations, $p(x_t|z_{0:t})$, also known as the belief $\mathcal{B}(x_t)$, is obtained using the Bayesian rule:

\begin{equation}\label{obs-eq}
\mathcal{B}(x_t) = p(x_t|z_{0:t}) = \frac{p(z_t|x_{t})\times p(x_{t}|z_{0:t-1})}{p(z_{t}|z_{0:t-1})} \propto p(z_t|x_{t}) p(x_{t}|z_{0:t-1}).
\end{equation}

By substituting $p(x_t|z_{0:t-1}) = \sum_{x_{t-1} \in \mathcal{X}} p(x_t|x_{t-1})p(x_{t-1}|z_{0:t-1})$ in the above equation (considering Markovian properties), the Bayes filter algorithm can be used to obtain a recursive update rule:
\begin{equation}\label{belief-eq}
\mathcal{B}(x_t) = p(z_t|x_{t})\sum_{x_{t-1} \in \mathcal{X}} p(x_t|x_{t-1}){B}(x_{t-1}),
\end{equation}
where, $p(x_t|x_{t-1})$ is the transition probability found similarly as in \cite{whitney2016interpreting}, 
\begin{equation}\label{transition-eq}
p(x_t|x_{t-1}) =\begin{cases}
c, & \text{if $x_t = x_{t-1}$}.\\
    \frac{1-c}{|\mathcal{X}|-1}, & \text{otherwise}.
  \end{cases},
\end{equation}
where $c$ is a constant value. 

The observation model calculates the probability of the observation given the state. Each observation is a set of the user's head movement, hands pointing, and speech, $<h, l, r, s>$ where:
\begin{itemize}
	\item $h$ represents a 3D vector of roll, pitch and yaw angles of the head.
    \item $l$ represents a 3D vector as the  direction of the participant's left  index finger.
    \item $r$ represents a 3D vector as the  direction of the participant's right  index finger.
    \item $s$ represents a list of words uttered by the participant.
\end{itemize}

More formally, the observation model looks as follows:

\begin{equation}
p(z_t|x_{t}) = p(h, l, r, s|x_t).
\end{equation}

We factor the expression by assuming that each observation is conditionally independent of the others given the target object. In other words,  if we know the intended target object, 
knowledge about e.g., right hand pointing does not provide any further information about the head movements. This results in the following factorization:

\begin{equation}
p(z_t|x_{t}) = p(h_t | x_t) \times p(l_t | x_t) \times p(r_t | x_t) \times p(s_t | x_t).
\end{equation}

In the following, we discuss how the above likelihoods can be modelled in our proposed approach. 

\subsubsection{Head Movement}\label{head}

We first learn a model p $\leftarrow f_h (h)$ that maps an angular position of the participant head ($h$) into a 2D position on the table ($p$) where the participant is looking at.
Following the guidelines of the device \cite{harezlak2014towards}, we calibrated it as an eye-tracker by training a Support Vector Regression (SVR) \cite{smola2004tutorial} with a RBF kernel (C=10, gamma=5) on 14 known points on the table. Participants were asked to look at each point for a duration of 1950 ms out of which the first 700 ms period was ignored. 
This calibration process results in $\pm4.85$ cm gaze positioning error. 



Similar to the earlier study \cite{whitney2016interpreting}, we assign distributions over different head angular positions according to the distance between the corresponding gaze location and the target object location, i.e., 

\begin{equation} \label{head-eq}
p(h_t|x_{t} = i) \propto \exp{[-(f_h(h_t)- p_i)^T\Sigma_h(f_h(h_t) - p_{x_{t}})]},
\end{equation}
where, $p_{i}$ is the position of the $i_{th}$ object on the table, and $\Sigma_h$ is a diagonal co-variance matrix with trainable parameters.

\subsubsection{Hand gestures}

Similarly, two separate SVM models are trained to map the directions of the left ($p \leftarrow f_l(l)$) and right ($p \leftarrow f_r(r)$) hands to the corresponding 2D positions on the table. 
Pointing detection is made with the help of LeapMotion device. 
The same expression as in eq.~\ref{head-eq} is used to assign probability distributions over left $l_t$ and right $r_t$ hand pointing directions conditioned on the target object $x_t$.


\subsubsection{Speech}

First, we use speech recognition to convert audio to text and then perform keywords dictionary-based classification to identify the following speech events: 

\begin{itemize}
    \item attribute - adjectives that describes size, shape or colour of a Lego block from the workspace, e.i. \textit{red, large, cylinder}, etc.
    \item deictic - i.e. \textit{here, there, this, that}, etc.
    \item other - any other word that is not included into the previous two categories
\end{itemize}

As an extra speech event, the beginning of a verbal request is detected from the audio directly. These specific classes  were inferred from the initial data collection. The highest correlation was shown between them and events in other modalities.

After detecting speech events, we represent it with a unigram model. Namely, we take each word $w$ in a given speech input $s$ and calculate the probability that, given state $x_t$, that word would have been spoken.

\begin{equation}
p(s|x) = \displaystyle\prod_{w\in s} p(w|x)
\end{equation}


\subsection{Temporal Priors}

The main hypothesis of this paper, is whether incorporating the knowledge of temporal correlations between high-level events in the input modalities can help the robot to better understand the intentions of a human while the person is referring to something.  
In order to validate this hypothesis, we propose to use temporal conditional probabilities to represent the observation likelihood introduced in eq.~\ref{obs-eq}. This yields,

\begin{multline}\label{obs-temp-eq}
p(s_t, h_t, \Delta T_h, l_t, \Delta T_l, r_t, \Delta T_r |x_t) =\\ = p(s_t|x_t) \times p(h_t|x_t) \times p(l_t|x_t) \times p(r_t|x_t) \times p(\Delta T_h,\Delta T_l,\Delta T_r|x_t),
\end{multline}
where $\Delta T_h = T_s - T_h$, i.e., the time difference between the speech and head movement events. Similarly,  $\Delta T_l = T_s - T_l$ and $\Delta T_r = T_s - T_r$. We used $T_s$ as the time reference, since it is less affected by noises compared to the other modalities. Furthermore, we assumed independence between, e.g., the current value of the head position and its event time. However, the time differences between the events are highly correlated.






In this paper, multivariate Gaussian Mixture Models (GMMs) are used to represent the PDF of $p(\Delta T_h,\Delta T_l,\Delta T_r|x_t)$. We assume that modalities are temporally dependent; thus, the co-varience matrix is learned alongside kernels' means. We train GMMs with an Expectation Maximisation (EM) algorithm. Online adaptation of the model is performed via Maximum A Posteriori (MAP) estimation approach, as in Reynolds' work \cite{reynolds2000speaker}, due to a limited number of samples that we are able to collect during the interaction.






\section{Data Collection}

In order to test our hypothesis of existing temporal dependencies in behaviour patterns, we collected data of people interacting with a robot controlled by a wizard of Oz. 




Our previous experiments showed that human behaviour varies dramatically in human-human interaction versus human-robot interaction. For instance, if the person has a human partner, they were more prone to use gestures and look at their partner. On the other hand, with a robot partner, participants favoured other modalities, like speech and exaggerated head movements. Moreover, robots can interpret some modalities with more precision than humans do. For humans, it is easier to understand where the person is pointing at than where they are looking. For robots using head movement tracking sensors, this modality becomes much more precise and easy to interpret than hand gestures. Our goal is to recreate the data collection scenario as close as possible to the target settings of the real human-robot interaction we plan our robot to operate in. 

\begin{figure}
\centering
\includegraphics[width=0.7\columnwidth,scale=0.5]{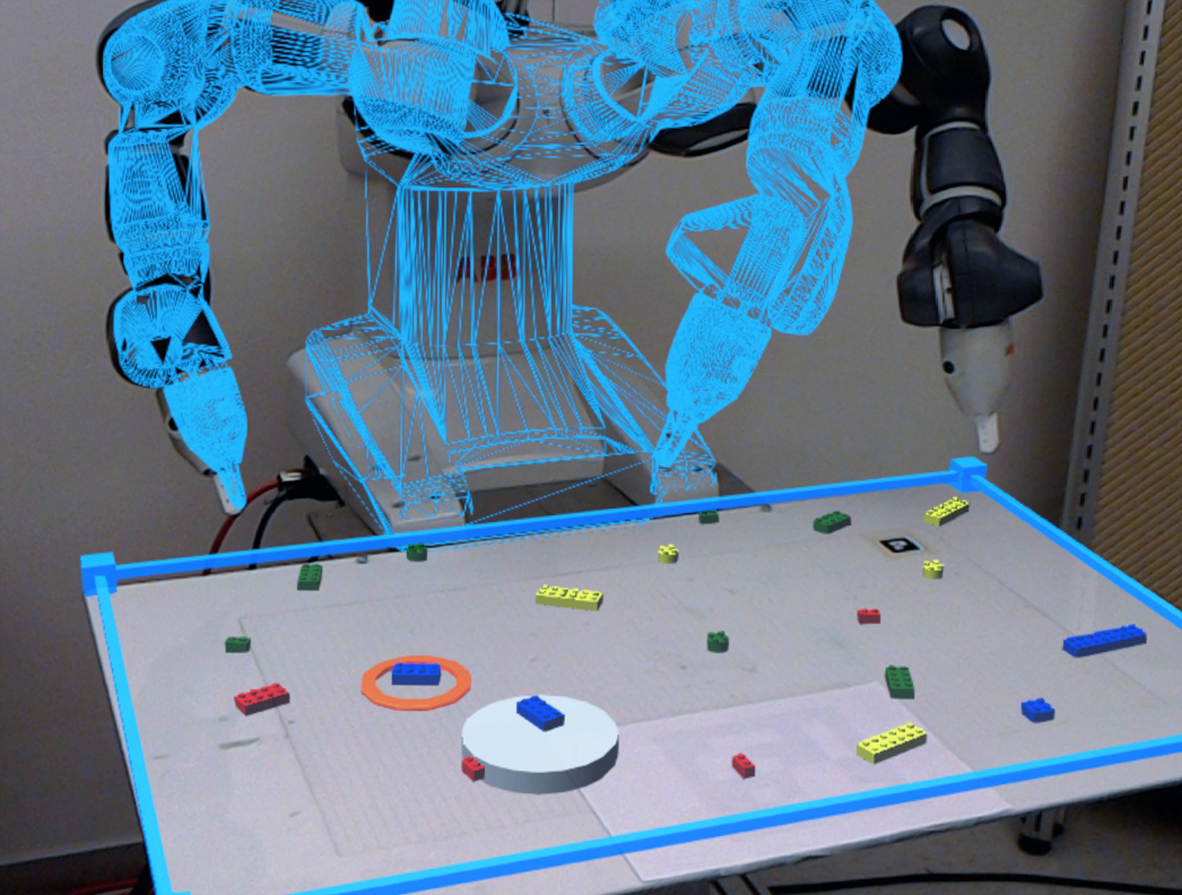}
\caption{First person view of the mixed reality interface. The participant is instructed to request an Lego block, that has an augmented orange circle around it. Robot's guess is indicated as a white cylinder. The virtual robot shows the future trajectory of a picking up motion before the participant confirms whether the object was disambiguated correctly. }
\label{fig:mr}
\end{figure}


\subsection{Scenario}
In our study, participants are instructed though a projection in a mixed reality headset to request Lego blocks (see Figure \ref{fig:mr}) from the robot in an ambiguous environment, i.e., there are several blocks of the same colour and shape. Thus, it is impossible to disambiguate a human request only from speech and the interpretation of other modalities is necessary. Mixed reality was chosen as the way to convey robot's current belief and augment additional information on the shared workspace, based on the results of our previous human study \cite{sibirtseva2018comparison}. A human wizard interprets the human requests by looking at what a robot would be able to sense and tries to infer the intended object. When the human participant acknowledges that the robot understood which object the participant meant, the mixed reality headset suggests the next object for the participant to describe to the wizard. The wizard's interface contains data from all the sensors. Namely,

\begin{itemize}
\item 3D position and rotation of the participant's head tracked by the headset\footnote{https://www.microsoft.com/en-us/hololens} and updated at frequency 60Hz;
\item Projection from the centre of their head on the table;
\item 3D coordinates of both hands, a projection from the index finger on the table when pointing occurs. The original frame rate of the sensor is 120Hz but to align data streams of head and hand tracking, we record only each second frame, resulting in 60Hz frequency. Tracking is performed by the Leap Motion sensor\footnote{https://www.leapmotion.com/};
\item Speech recognition represented as text, acquired from using Microsoft Speech Platform\footnote{https://msdn.microsoft.com/en-us/library/hh361572(v=office.14).aspx}. Speech data is recorded every time the dictation hypothesis is updated. We don't wait for the utterance to be completed, and instead employ a riskier but also faster approach.
\end{itemize}

\begin{figure}
\centering
\begin{overpic}[width=0.7\columnwidth,scale=0.5]{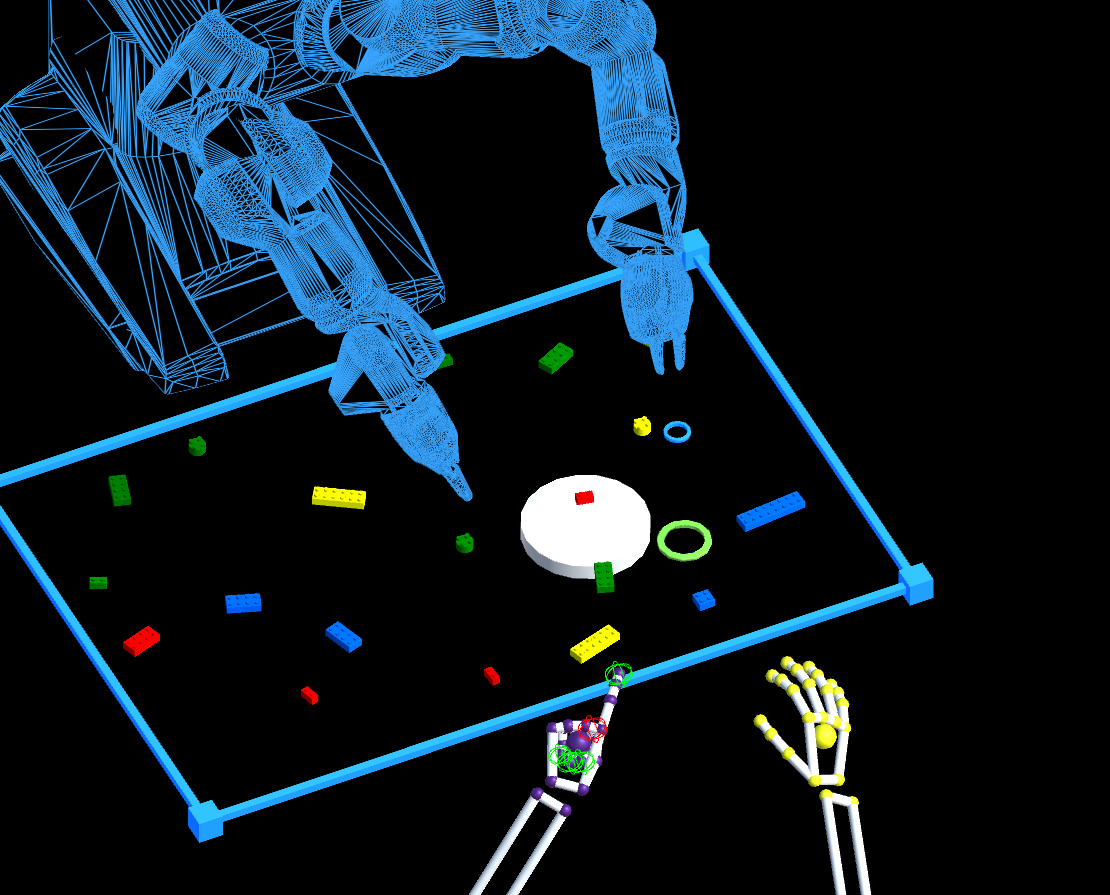}
\put (63,68) {\color{white}\textsf{please give me this}}
\put (63,63) {\color{white}\textsf{um... small block}}
\end{overpic}
\caption{Wizard interface with wizard's guess visualised as a white cylinder around an object. Multimodal input is represent as (a) text from speech recognition; (b) white rectangle being participant's head position and rotation, while a blue circle on the table surface is a projected vector from the centre of the head; (c) skeleton of a hand and a projected position from its finger on the table plane as a green circle. }
\label{fig:wizard}
\end{figure}






\subsection{Participants}

Subjects were recruited using mailing lists and flyers on the university campus. A total of 30 subjects (16 female, 14 male), ages between 23 and 34 ($M=27.7$), participated in the data collection. All participants had to meet the following requirements: to be fluent in English, not require glasses to see objects 1.5-2 away from them (due to the mixed reality head-mounted display) and not have any colour vision deficiencies. 
In general, participants indicated their experience with digital technology as $M=1.3$ on a scale from 5 to 1 (where 1 denotes \enquote{very highly}). Moreover, $57\%$ had some experience with virtual reality and only $10\%$ tried augmented/mixed reality head-mounted displays before. 

\subsection{Dataset}

Each participant made a total of 20 fetching requests. The time of each request was not fixed; the start time of the request is considered to be the moment a participant was shown a new object in the Mixed Reality interface, and the request was considered resolved when a participant verbally confirmed the robot's guess. Thus, enabling the data collection to proceed to the next randomly selected object and marking the current timestamp as the end of the request. Overall, 600 requests were collected with a total duration of 429 minutes of uninterrupted recording. Each request consisted of multiple datapoints with the following fields: a timestamp, a 3D vector of \textit{head position}, a boolean variable representing whether the current head movement is a \textit{fixation}, a 3D vector of each hand index \textit{finger positions}, a boolean variable indicating whether the current gesture was \textit{pointing}, the \textit{text of the verbal request} so far from speech recognition, the current \textit{target object}, and current \textit{wizard's guess} of the target object. The final dataset contains $N=30705$ datapoints.

\section{Results}

This section presents our findings on the questions \textbf{Q1} - \textbf{Q3}.

\subsubsection{\textit{Q1:}} \textbf{\textit{Do common temporal patterns emerge in participants' behaviour during the fetching request task?}}

A pre-processing step is performed before training temporal priors encoded as GMMs. All fixations from both head movements and gestures are labelled as intentional or accidental. By intentional, we imply a fixation on the target object. All the other fixations are considered accidental, or noise. In a request, fixations on the minimum distance from the ground truth target object are labelled as intended. Moreover, time intervals of head fixations and pointing gestures are computed relative to the events in the speech modality. As a result, the training data set consists of time intervals and labels of head fixations, gestures and types of the corresponding event in the speech modality. Finally, through leave-one-participant-out cross validation, we train GMMs temporal priors on the training dataset. Analysis of the GMMs densest regions discovers three most common temporal patterns in participants' behaviour, namely:

\begin{itemize}
    \item \textbf{P1} Head fixation + beginning of the verbal request
    \item \textbf{P2} Head fixation + deictic keyword + pointing gesture
    \item \textbf{P3} Object attribute keyword + head fixation
\end{itemize}

\begin{figure}
\centering
\includegraphics[width=\columnwidth,scale=1]{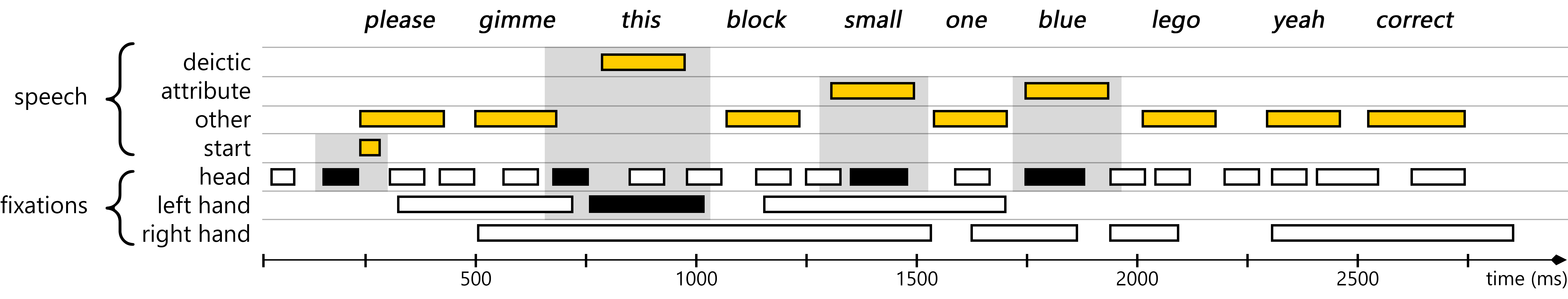}
\caption{A timeline of events in each modality during request 3 from participant 27. The top line is speech recognition of participant's request. Black rectangles represent fixations on the target object, white on any other. Grey rectangles indicate common behaviour patters. }
\label{fig:temporal}
\end{figure}

Let's observe how these patterns appear in the human-robot interaction during a fetching request. On the Figure \ref{fig:temporal}, we visualise a timeline of events in each modality during one of the participant's request. This request contains all three of the common patterns (highlighted with rectangles). In the beginning of the request, the participant, firstly, fixates his/her head on an unrelated to the request object. Then, he starts to verbalise his request with words \enquote{please gimme...}. Just before the beginning of the utterance, the first common pattern occurs (\textbf{P1}) - the participant fixates on the target object. After that several unintentional head fixations are detected alongside with the left hand pointing gesture in an uninformative direction. At the timestamp 800 ms the second pattern (\textbf{P2}) is detected - deictic keyword \enquote{this} with an intended head fixation and the left hand pointing at the target object. Later on, at timestamps 1300 ms and 1800 ms we can observe \textbf{P3}. The participant clarifies his request by saying an attribute key-word \enquote{small}and \enquote{blue} simultaneously with head fixations on the target object.

\subsubsection{\textit{Q2:}} \textbf{\textit{If we encode these patterns as temporal priors, will they be helpful in inferring the intended object from multimodal referring expressions?}}

To answer this question, we compare performance of Bayesian filter with (\textbf{BF+TP}) and without (\textbf{BF}) temporal priors through leave-one-participant-out cross validation. Bayesian filter without temporal priors (\textbf{BF}) is considered as a control condition. Our evaluation consists of two measures - \textbf{accuracy} (\%) and decision making \textbf{time} (sec). Accuracy is measured as a ratio of correctly disambiguated target objects to the total number of requests. Decision making time is computed from the moment a person started speaking and until the robot makes a final decision which object is the target. 


There are two possible scenarios of how a model can make a decision. 

\begin{itemize}
    \item\textit{ Voluntary decision}. We define a decision making line as 85\%, a commonly used value for such scenarios \cite{whitney2016interpreting}. This means that if probability of an object being the target reaches 85\% or higher, given all previous multimodal observations, then the model selects it as a target object.
    \item \textit{Forced decision}. If during a request no object's probability to be the target one crossed the decision line and there are no datapoints left in this request, then the object with the highest probability is chosen the target.
\end{itemize}

Models have to make forced decisions due to the way we collected the dataset. A human wizard was imitating a robot during the data collection process; therefore, a request was considered resolved when the wizard guessed the correct target object. However, our models have a limited understanding of multimodal human behaviour and are not as sophisticated as a human wizard reasoning. 

To find out what causes the main issues for the models, compared to the human wizard, we analysed only cases where models had to make forced decisions. We manually checked video recordings of the interactions and discovered that the majority of such cases contained utterances such as \enquote{to the left}, \enquote{in the corner}, \enquote{he same as the previous one} A human is able to infer much more information from such phrases, while our speech system is only corpus-based. This bottleneck can be addressed by implementing a more sophisticated natural language recognition system, for instance, BERT \cite{devlin2018bert}.

For models' evaluation we take into account both voluntary and forced decisions.

Two types of evaluation is performed. In the first case, only the first attempt of the model to make a decision is considered. In the second, though, we employ the same way of interaction as during the data collection phase. The model can make several attempts to guess the target object while there is still data left in the request. After each guess it gets feedback whether the guess is correct or not. If it is incorrect, the model excludes the previous object from the possible objects set and proceeds disambiguation participant's request as before. For the multiple attempts case, the decision making time is measured from the beginning and until the guess is either correct or the request is over and the model is forced to make the final guess. 


We performed a repeated measurement one-tailed t-test to test significance of our results on the 95\% interval. According to the Table \ref{tab:resultstab1}, \textbf{BF+TP} dramatically $(p<0.00)$ decreases decision making time from $24.99 \pm 7.94$ sec to $15.32 \pm 3.08$ sec. Accuracy of the Bayesian Filter with Temporal Priors $(M = 68.45, SE =5.73)$ is also significantly $(p<0.00)$ higher than without $(M = 55.83, SE = 12.01)$. 

In the multiple attempts case (Table \ref{tab:resultstab2}), the tendency of \textbf{BF+TP} $(M_{time} = 18.85, SE_{time} = 3.73, M_{acc} = 86.22, SE_{acc} = 4.34)$ outperforming BF on both measurements is even more evident. Time and accuracy of \textbf{BF} does not significantly $(p>0.05)$ change from the first attempt. Number of attempts per request gives us insight into why this is the case. For BF, number of attempts is nearly one per request $(M = 1.14)$, while \textbf{BF+TP} can make $2.58$ on average. An explanation to this can be drawn from the Table \ref{tab:resultstab2}, specifically the decision making time. \textbf{BF} takes approximately $1.6$ times more to make the first decision and it does not have enough time left of the request to make an accurate guess on the second attempt. In a multiple attempt scenario \textbf{BF+TP} model can potentially make more guesses on the same amount of data, while being more accurate than the control condition.

Therefore, we can conclude that temporal priors have a significant positive influence on both decision making time and accuracy for the both evaluation scenarios.

\begin{table}
\centering
\caption{ Models evaluation on the first attempt at disambiguating a multimodal referring expression}
\label{tab:resultstab1}
\begin{tabular}{|l|c|c|} \hline
Model & Time (sec) & Accuracy\\ \hline
BF & 24.99 $\pm$ 7.94 & 55.83\% $\pm$ 12.01\%\\ 
BF+TP & 15.32 $\pm$ 3.08 & 68.45\% $\pm$ 5.73\%\\ 
BF+TP+OA & \textbf{13.41 $\pm$ 2.93} & \textbf{76.58\% $\pm$ 5.65\%}\\ \hline
\end{tabular}
\end{table}

\begin{table}
\centering
\caption{ Models evaluation on multiple number of attempts at disambiguating a multimodal referring expression}
\label{tab:resultstab2}
\begin{tabular}{|l|c|c|c|} \hline
Model & Time (sec) & Accuracy & \# Attempts\\ \hline
BF & 25.50 $\pm$ 8.11 & 57.38\%$ \pm$ 12.24\% & 1.14\\ 
BF+TP & 18.85 $\pm$ 3.73 & 86.22\% $\pm$ 4.34\% & 2.58\\ 
BF+TP+OA & 18.92 $\pm$ 3.70 & \textbf{89.16\% $\pm$ 4.28\%} & 2.63\\ \hline
\end{tabular}
\end{table}

\subsubsection{\textit{Q3:}} \textbf{\textit{Are temporal patterns common across most participants or person-dependent?}}

Our approach to \textbf{Q3} is to evaluate what is the effect of online adaptation (\textbf{BF+TP+OA}) on the decision making time and accuracy versus no participant-based adaptation (\textbf{BF+TP}). The model with online adaptation of temporal priors is performed in the same fashion as in the previous section, through leave-one-participant-out cross validation. However, we iteratively update the GMMs temporal priors by feeding them datapoints from the previously resolved request. The following request disambiguation is made with the refitted with all the previous requests model. This implies that with time, the GMMs become more fitted to the temporal patterns of this particular participant. The first requests of each participant in both models with and without adaptation are based on the same temporal priors GMMs. 

Our results show that while \textbf{BF+TP+OA} accuracy $(M = 76.58, SE = 5.65)$ increases significantly $(p=0.04)$ in comparison to the temporal priors without online adaption during the first attempt (Table \ref{tab:resultstab1}). Decision making time $(M = 13.41, SE = 2.93)$ , even though has a decreasing trend, is not statistically significant (p=0.13). 

For the multiple attempts case (Table \ref{tab:resultstab2}), adaptation results also show the best accuracy $(M = 89.16, SE = 4.28)$ out of the evaluated models. In regards of decision making time $(M = 18.92, SE = 4.28)$ and number of attempts, there was no statistically significant difference found between models \textbf{BF+TP} and \textbf{BF+TP+OA}.

We can reason that adaptation has a positive effect on model's accuracy, slightly adjusting temporal priors for each participant. The structure of the common patterns stays mostly the same between participants, while the Gaussian Mixtures shift to accommodate to the different timing of each participant individually.

\section{Conclusion}

In this paper, we explored temporal dependencies in multimodal human-robot interaction and developed a Bayesian-based model to evaluate our hypothesis. We developed a system in Mixed Reality to efficiently collect data of humans interacting with a robot in a fetching scenario. As our results showed, taking temporal dependencies between high-level events in all input modalities (i.e. fixations in head movements, key words in speech, etc.) increases the model's speed and accuracy of the person's intention predictions. Moreover, we tested how online adaptation influences results of the prediction and found out that, while both speed and accuracy increase, the change is not as dramatic as between using a Bayesian filter with or without temporal priors. Thus, we came to a conclusion that common temporal patterns exist in human behaviour during referencing objects and have a significant impact on the intention prediction. We encoded temporal priors as a Gaussian Mixture Model and used it with the Bayesian filter to compute probabilities of objects being the target ones. 

The next step for this project is to test how scalable our approach is to more complex tasks. Our initial motivation to explore temporal dependencies comes from our previous work \cite{kontogiorgos2018multimodal}, where participants where building furniture together. The main challenge there came from the nosiness of input modalities. And the more complex the interaction, the nosier participant's behaviour is. In other words, participants are less distracted and more focused on the task during simple interactions, such as fetching requests. We see the potential benefits of employing temporal priors to tackle nosiness in the more complex interactions. 

Another direction will be to add more modalities and explore how they can be represented as high-level events and encoded into temporal behaviour patterns. For instance, body posture and gaze tracking. A more nuanced, not key-words based, approach to natural language understanding can also enrich the possibilities for diverse interactions.  

And finally, in the future we would like to focus more on the deep reinforcement learning approaches to multimodal human-interaction. So far our study was necessary for gaining a deeper understanding of human behaviour and multimodal data. However, we want to move away from feature engineering and formulate our human-robot interaction scenario as a deep reinforcement learning problem. Recent studies in HRI showed impressive results in employing deep reinforcement learning for various applications \cite{mnih2015human,qureshi2016robot,lathuiliere2018neural}. The main challenge for deep learning approaches is the lack of training data from human studies but we plan to tackle this problem using our current Bayesian-based model to simulate human behaviour data as a prior for the deep reinforcement learning model.

\section{Acknowledgements}

This work is supported by the SSF (Swedish Foundation for Strategic Research) projects COIN.

\bibliographystyle{splncs_srt}
\bibliography{bibliography}

\end{document}